\documentclass{bioinfo}
\copyrightyear{2015} \pubyear{2015}

\access{Advance Access Publication Date: Day Month Year}
\appnotes{Manuscript Category}

\usepackage{url}

\begin{document}
\firstpage{1}

\newtheorem{theorem}{Theorem}
\newtheorem{corollary}{Corollary}
\newtheorem{lemma}{Lemma}
\newtheorem{proposition}{Proposition}
\newtheorem{definition}{Definition}

\newtheorem{example}{Example}
\newtheorem{remark}{Remark}

\newcommand{\TODO}[1]{\begingroup\color{red}#1\endgroup}

\subtitle{Structural bioinformatics}

\title[An efficient dual sampling algorithm with Hamming distance filtration]{An efficient dual sampling algorithm with
       Hamming distance filtration}
\author[Fenix Huang \textit{et~al}.]
       {Fenix W. Huang\,$^{\text{\sfb 1}}$, Qijun He\,$^{\text{\sfb 1}}$, Christopher Barrett\,$^{\text{\sfb 1,4}}$
         and Christian M. Reidys\,$^{\text{\sfb 1,2,3}*}$}
\address{
$^{\text{\sf 1}}$ Biocomplexity Institute of Virginia Tech, Blacksburg, VA, USA.
$^{\text{\sf 2}}$ Department of Mathematics, Virginia Tech, Blacksburg, VA, USA.
$^{\text{\sf 3}}$ Thermo Fisher Scientific Fellow in Advanced Systems for Information Biology.
$^{\text{\sf 4}}$ Department of Computer Science, Virginia Tech, Blacksburg, VA, USA.
}

\corresp{$^\ast$To whom correspondence should be adressed.}

\history{Received on XXXXX; revised on XXXXX; accepted on XXXXX}

\editor{Associate Editor: XXXXXXX}

\abstract{
  {\bf Motivation:}
  Recently, a framework considering RNA sequences and their RNA secondary structures as pairs, led to some
  information-theoretic perspectives on how the semantics encoded in RNA sequences can be inferred. 
In this context the pairing arises naturally from the energy model of RNA secondary structures. Fixing the sequence in the pairing produces the RNA energy landscape, whose partition function was discovered by McCaskill. Dually, fixing the structure induces the energy landscape of sequences.
%In this context a pairing of sequences and structures arises naturally, whose fiber with respect to fixed sequences is the partition function of secondary structures, discovered by McCaskill. Dually, the pairing contains fibers with respect to fixed structures, i.e.,~partition functions of sequences.
The latter has been considered for designing more efficient inverse folding algorithms. \\
    {\bf Results:}
    We present here the Hamming distance filtered, dual partition function, together with a Boltzmann sampler using novel dynamic programming
    routines for the loop-based energy model. The time complexity of the algorithm is
  $O(h^2n)$, where $h,n$ are Hamming distance and sequence length, respectively, reducing the time complexity of samplers,
  reported in the literature by $O(n^2)$.
  We then present two applications, the first being in the context of the evolution of natural sequence-structure
  pairs of microRNAs and the second constructing neutral paths.
  The former studies the inverse fold rate (IFR) of sequence-structure pairs, filtered by Hamming distance, observing 
  that such pairs evolve towards higher levels of robustness, i.e.,~increasing IFR. 
  %The second is the structural diversity produced by Hamming distance filtered sequences: here we consider scalar order parameters as well as constructing certain base-pairing probability matrices of the structures folded by the sampled sequences.   We observe that natural sequence-structure pairs evolve towards higher structural diversity.
  The latter is an algorithm that construct neutral paths: given two
  sequences in a neutral network, we employ the sampler in order to construct short paths connecting them, consisting
  of sequences all contained in the neutral network. \\
  \textbf{Availability:}
The source code is freely available at {http://staff.vbi.vt.edu/fenixh/HamSampler.zip}\\
\textbf{Contact:} \href{duckcr@bi.vt.edu}{duckcr@bi.vt.edu}\\
\textbf{Supplementary information:}
Supplementary material containing additional data tables are available at \textit{Bioinformatics} online.}

\maketitle

%Purpose of the paper: 
%1. algorithm, sample sequence with Hamming distance filtration
%2. why Hamming distance is important 
%a. evolutionary accessible. 
%b. higher IFR, more focus on the reference sequence. 
%3. Result, why is useful, what can be applied for. 

\section{Introduction}\label{S:Into}

\begin{figure}[h]
\begin{center}
\includegraphics[width=0.8\columnwidth]{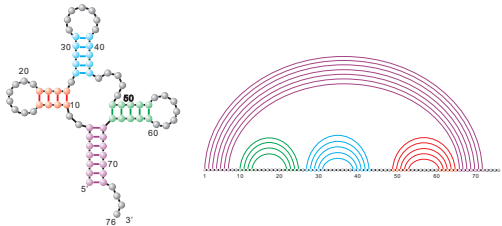}
\end{center}
\caption{\small tRNA: secondary structure and diagram representation. 
}
\label{F:RNAp}
\end{figure}

Ribonucleic acid (RNA) is a polymeric molecule essential in various biological roles. RNA consists of a single strand of
nucleotides ({\bf A, C, G, U}) that can fold and bond to itself through base-pairings. At first, RNA was regarded as a simple
messenger - the conveyor of genetic information from its repository in DNA to the ribosomes.  Over the last several
decades, however, researchers have discovered an increasing number of important roles for RNA. RNAs have been found to
have catalytic activities, to participate in processing of messenger RNAs, to help maintain the telomers of eukaryotic
chromosomes, and to influence gene expression in multiple ways \citep{darnell2011rna, breaker1996engineered, serganov2007ribozymes, breaker1994inventing}. The specific shape into which RNAs fold plays a major role in their function, which makes RNA
folding of prime interest to scientists. An understanding of RNA's three-dimensional structure will allow a greater
understanding of RNA function. However, obtaining these three-dimensional structure through crystallization is often
costly and time consuming. Accordingly coarse grained RNA structures are considered, the most prominent of which being
RNA secondary structures. The latter are contact structures with noncrossing arcs when presented as a diagram, 
see Fig.~\ref{F:RNAp}.

The key feature of RNA secondary structures is that they can be inductively constructed\footnote{considered as fatgraphs
of genus zero they are the Poincar\'{e} dual of planar trees} \citep{stein1978class}.
Waterman {\it et al.} \citep{Waterman:78a, stein1978class, Nussinov:78, Kleitman:70} studied the combinatorics and
folding of RNA secondary structures. 
%Their diagrams are labeled graphs over the vertex set $[n]=\{1, \dots, n\}$, presented by drawing the vertices on a horizontal line and noncrossing arcs in the upper half-plane. Vertices and arcs correspond to the nucleotides {\bf A}, {\bf G}, {\bf U} and {\bf C} and Watson-Crick ({\bf A-U}, {\bf G-C}) and wobble ({\bf U-G}) base-pairs, respectively.
The noncrossing arcs of RNA secondary structures allow for a recursive build: let $S_2(n)$ denotes the number of RNA
secondary structures over $n$ nucleotides then we have \citep{Waterman:78a}: $S_2(n)=S_2(n-1)+
\sum_{j=0}^{n-3}S_2(n-2-j)S_2(j)$, where $S_2(n)=1$ for $0\le n\le 2$. The recursion forms the basis for more than three
decades of research resulting in what can be called the dynamic programming (DP) paradigm. The
DP paradigm allows one to compute minimum free energy (MFE) structure in $O(n^3)$ time and $O(n^2)$ space.
Implementations of these DP folding algorithms are mfold and {\it ViennaRNA} \citep{Zuker:81, Hofacker:94a}, employing
the energy values derived in \citep{Mathews:99, Turner:10}. The so called inverse folding,
i.e.,~identifying sequences that realize a given structure as MFE-structure, has been studied in \citep{Hofacker:94a, Busch:06}.

\vspace{-10pt}

MFE folding naturally induces a genotype-phenotype (sequence to structure) map, in which the preimage of a structure is
called the \emph{neutral network}. Neutral networks are closely related to the \emph{neutral theory} of Motoo Kimura
\citep{Kimura:68}, which stipulates that evolution is driven by mutations that do not change the phenotype. The properties
of neutral networks as subsets of sequences in sequence space allow one to study how genotypes evolve. Neutral networks have been
studied theoretically via random graph theory \citep{Reidys:97}, in the context of the molecular
quasispecies \citep{Reidys:97b} and by exhaustive enumeration\citep{Gruner:96a, Gobel:00}. A neutral network
represents the set of all inverse folding solutions of a fixed structure. Graph properties, like for instance, size, density
and connectivity are of crucial functionality in molecular evolution. Clearly, a vast, extended neutral network is more
accessible than small, localized one and on a connected and dense neutral network, neutral evolution can easily be
facilitated via point- and pair-mutations. On such a network, a population of RNA sequences can explore sequence space via gradual
genotypic changes while maintaining its phenotype. 

However, there is more to sequences and structures than MFE-folding: certain RNA sequences exhibit multiple, distinctively
different, stable configurations \citep{baumstark1997viroid, schultes2000one}, as for example, riboswitches
\citep{Sergano:07,mandal2004adenine}. Recently \citep{reza2017driftwalks} evolutionary trajectories, so called {\it drift walks} have been
considered that are obtained by either neutral evolution or switching between a multiplicity of MFE-structures present at a
fixed sequence.
Such sequences indicate that is may not suffice to consider merely the MFE-structure, but rather to broaden the scope to the
entire RNA energy landscape. Energy landscapes of sequences, i.e.,~the spectrum of free energies of the associated secondary structures of a fixed
sequence have been studied in physics, chemistry, and biochemistry, and play a key role in understanding the dynamics of both
RNA and protein folding \citep{dill1997levinthal, onuchic1997theory, martinez1984rna, wolfinger2004efficient}. 

In \citep{McCaskill:90}, McCaskill observed that the tropicalization of the DP routine that computes the MFE-structure
produced the partition function of structures for a given sequence. This allows one to study statistical features, as, for
instance, base-pairing probabilities of RNA energy landscapes by means of Boltzmann sampling \citep{Tacker:96a, Ding:03},
enhancing structure prediction \citep{Ding:03, Bernhart:06, Heitsch:14}.
Aside from global features, local features are being studied: for instance, local minima of the energy landscape,
i.e.,~`energy traps' are crucial to the understanding of folding dynamics since they represent the metastable configurations
\citep{chen2000rna, tinoco1999rna}. Statistical features of constrained energy landscapes, corresponding to conditional
distributions can also be Boltzmann sampled \citep{Hofacker:94a, freyhult2007boltzmann, lorenz20092d}.

Accordingly, the partition function is tantamount to computing the probability space of structures that a fixed sequence is
compatible with. This gives rise to consider the pairing \citep{Barrett:17}:
\begin{equation}
  \eta \colon \mathcal{N}^n \times \mathcal{S}_n \longrightarrow \mathbb{R},
\end{equation}
which maps a fixed sequence-structure pair into its free energy. Here $\mathcal{N}^n$ and $\mathcal{S}_n$ denote the space
of sequences, $\sigma$, and the space of secondary structures, $S$, respectively. 
The pairing illuminates the symmetry between sequences and structures, suggesting to 
consider the ``dual'' of RNA energy landscape, i.e.,~the spectrum of free energies of sequences with 
respect to a fixed structure. 
%consider the ``dual'' of McCaskill's partition function, i.e.,~the partition function of all sequences that are compatible with a fixed structure. 
This dual has been employed for designing more efficient inverse folding algorithms: \citep{Busch:06} discovers
that using the MFE sequence of a fixed structure as starting point for the inverse folding, significantly accelerates the
algorithm. In other words, the global minimum of the RNA dual energy landscape is typically very close in sequence space
to the corresponding neutral network. This line of work motivated the use of the dual RNA energy landscape\footnote{by
sampling} in inverse folding algorithms \citep{levin2012global,Clote:16}.
Recently, \citep{Barrett:17} proposed a framework considering RNA sequences and their RNA secondary structures simultaneously,
as pairs. The RNA dual energy landscape in this context gives rise to an information theoretic framework for RNA sequences.

In practice, the exhaustive exploration of the dual RNA energy landscape is not feasible, whence specific localizations,
for instance studying the point-mutant neighborhood of a natural RNA sequence \citep{borenstein2006direct, rodrigo2012describing}
have been studied.

To conduct a systematic and biologically meaningful study of the dual RNA energy landscape, we present in this
paper an efficient Boltzmann sampling algorithm with a Hamming distance filtration. This filtration facilitates
the analysis of Hamming classes of sequences in the dual RNA energy landscape, that would otherwise be
impossible to access, see Fig.~\ref{F:distribution}. Instead of being restricted to neighborhoods of point-mutants
\citep{borenstein2006direct, rodrigo2012describing}, we have now access to arbitrary Hamming classes. Such a dual
sampler has to our knowledge first been derived in \citep{levin2012global}. In fact, the sampler arises as the restriction
of \citep{waldispuhl2008efficient}, where the structure partition function of sets of sequences with fixed Hamming distance
is computed. As a result, its recursions over subintervals, that form the conceptual backbone, lead to a time complexity
of $O(h^2n^3)$, where $h$ and $n$ denote Hamming distance and sequence length, respectively. In contrast, the Boltzmann sampler
presented here is based on the loop-decomposition of the fixed structure and has a time complexity of $O(h^2n)$.

The paper is organized as follows: in Section 2, we discuss our sampling algorithms. In Section 3, we study two
application contexts of the dual Boltzmann sampler with the Hamming distance filtration. First we study the inverse fold rate as
a function of Hamming distance and then we employ our dual sampler in order to explicitly construct neutral paths in neutral networks.

%%%
%%%%%%%%%%%%%%%%%%%%%%%%%%%%%%%%%%%%%%%%%%%%%%%%%%%%%%%%%%%%%%%%%%%%%%%%%%%%%%%%%%%%%%%%%%%%%%%%%%%%%%%%%%%%%%%%%%%%%
%%%
\section{Methods}\label{S:Method}
%%%
%%%%%%%%%%%%%%%%%%%%%%%%%%%%%%%%%%%%%%%%%%%%%%%%%%%%%%%%%%%%%%%%%%%%%%%%%%%%%%%%%%%%%%%%%%%%%%%%%%%%%%%%%%%%%%%%%%%%%
%%%

%Briefly introduce what has been done, summary of our method, shinning point.
In \citep{Busch:06} a minimum free energy (MFE) sequence for a given structure is derived by means of dynamic programming (DP).
The algorithm facilitates the arc decomposition of a secondary structure \citep{Waterman:78a} computing a MFE sequence
recursively. In analogy to the partition function of structures, the dual partition function has been computed in
\citep{Clote:16, Barrett:17}, where in addition Boltzmann samplers were derived \citep{Clote:16, Barrett:17}.

In this section we introduce an algorithm refining the Boltzmann sampler in \citep{Barrett:17} that constructs RNA sequences from the
Boltzmann ensemble of a structure $S$, subject to a Hamming distance constraint\footnote{to a given reference sequence
  $\overline{\sigma}$, say}. The straightforward approach would be to run a rejection sampler based on the sampler introduced in
\citep{Clote:16, Barrett:17}. However, as we shall prove in Section~\ref{S:Result}, this would result in a rather inefficient algorithm.
Instead, we follow a different approach, introducing a new parameter $h$ associated to a subsequence, representing the Hamming distance.

%Preliminary, graph presentation of secondary structure (short), what is a loop, what energy model we are using, how to compute energy
Let us first recall the graph presentation of RNA secondary structures: RNA secondary structures can be represented as diagrams,
where vertices are drawn in a horizontal line and arcs in  the upper half-plane. In a diagram, vertices are presenting nucleotides
and arcs are presenting base-pairs, see Fig.~\ref{F:RNAp}. 
Vertices are labeled by $[n] = \{1, 2, \ldots, n\}$ from left to right, indicating the orientation of the backbone from the
$5'$-end to $3'$-end. A base-pair, denoted by $(i, j)$ is an arc connecting vertices labeled by $i$ and $j$. 
Two arcs $(i, j)$ and $(r, s)$ are called {\it crossing} if for $i<r$, $i<r<j<s$, holds. An RNA secondary structure contains
exclusively noncrossing arcs and thus induces the partial order: $(r,s) \prec (i,j)$ if and only if $i < r < s < j$.

%loops
The energy of a sequence-structure pair $\eta(\sigma, S)$ can be computed as the sum of the energy contributions of individual
base-pairs \citep{Nussinov:78}. A more elaborate model \citep{Mathews:99, Turner:10} evaluates the total free energy to be the
sum of from the energies of loops involving multiple base-pairs.
A loop $L$ in a secondary structure is a sequence of intervals $([a_i, b_i])_i$, $1\le i \le k$, 
where $(a_1, b_k)$, $(b_i, a_{i+1})$, $\forall 1\le i \le k-1$, are base-pairs. Since no crossing arcs are allowed, 
nucleotides in the interval $([a_i+1, b_i-1])_i$ are unpaired. In particular, for $k=1$, $L$ is called a hairpin loop
for $k=2$ either an interior loop, bulge loop or helix, depending on how many unpaired vertices are contained in the respective
intervals, and for $k\ge 3$, a multiloop. 
Note that the arc $(a_1, b_k)$ is the maximal arc of the loop, i.e., $(b_i, a_{i+1}) \prec (a_1, b_k)$ for all $1\le i \le k-1$, 
whence $L$ can be represented by $(a_1, b_k)$. The intersection of two distinct loops is either empty or consists of exactly
one base-pair.
Each base-pair is contained in exactly two loops and is maximal in exactly one of these two.
There is a particular loop, the {\it exterior} loop, consisting of all maximal arcs in a secondary structure.
As a matter of convention, we shall assume that any diagram is ``closed'' by the arc, $(0, n+1)$, referred to as its
rainbow and by convention, there are the two ``formal'' nucleotides $N_0,N_{n+1}$ associated with positions $0$ and $n+1$,
respectively.

%loop energy 
In Turner's energy model, the energy of a loop, $\eta(\sigma, L)$, is determined by its loop type 
(hairpin, interior loop, exterior loop or multiloop), the specific nucleotide composition of its base-pairs as well as a certain
number of unpaired bases contained in it. Those unpaired bases are typically adjacent to a base-pair.
%The latter are typically the bases associated with a base-pair and those adjacent to a base-pair. 
Accordingly, the energy of a sequence-structure pair equals the sum of the energies of all the associated loops, i.e., 
\begin{equation}\label{E:energy}
\eta(\sigma, S) = \sum_{L\in S} \eta(\sigma, L). 
\end{equation}

\begin{figure}[h]
\begin{center}
\includegraphics[width=0.9\columnwidth]{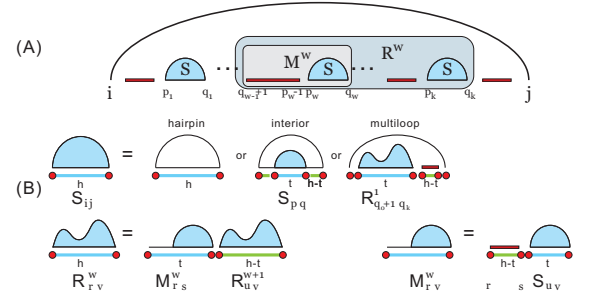}
\end{center}
\caption{\small (A): the substructures $S_{a,b}$, $M^w_{a,b}$ and $R^w_{a,b}$. 
(B) structural decomposition and Hamming distance distribution. 
} 
\label{F:ham_decom}
\end{figure}

%arc decomposition
A secondary structure can be decomposed by successively removing arcs from the outside to the insider (top to bottom), see
Fig.~\ref{F:ham_decom}. Since any base-pair is maximal in exactly one loop, removing a base-pair is tantamount to
removing its associated loop.

Viewing a secondary structure, $S$, as a diagram we observe that any interval $[i,j]$ induces a substructure containing
all arcs that have both endpoints contained in $[i,j]$ and denote such substructures by $X_{i,j}^S$. In case the interval
$[i,j]$ contains no arcs, we simply refer to the substructure $X_{i,j}^S$ again as an interval. Given $S$, the concatenation
of the two substructures $X_{i,j}^S \cup X_{j+1,k}^S$ is the substructure $X_{i,k}^S$. In the following we shall simply write
$X_{i,j}$ instead of $X^S_{i,j}$.

In particular, let $(i, j)$ be a base-pair, $L$ be the loop that is represented by $(i,j)$ and let $S_{i,j}$ be the
substructure for which $(i,j)$ is {\it the} maximal arc.
Suppose $(p_r, q_r)$, $1\le r\le k$ are base-pairs in $L$, different from $(i, j)$, removing the arc $(i,j)$ produces a
sequence of substructures $S_{p_1,q_1}, \ldots S_{p_k,q_k}$ as well as a sequence of intervals $[i+1, p_1-1], [q_1+1, p_2-1],
\ldots, [q_k+1, j-1]$.

Let $q_0 = i$, concatenating the interval $[q_{r-1}+1, p_r-1]$ with $S_{p_r,q_r}$ produces a substructure, which we denote by
$M^r_{q_{r-1}+1,q_r}$, $1\le r\le k$. Let $R_{q_{0}+1,q_k}^1$ be the substructure obtained by concatenating all
$M^r_{q_{r-1}+1,q_r}$ for $1\le r\le k$, i.e.,~$\bigcup_{r}M^r_{q_{r-1}+1,q_r}$. By construction, removing $(i,j)$ from $S_{i,j}$
generates $R_{q_0+1,q_k}^1\cup [q_k+1,j-1]$. 

Note that $R^1_{q_0+1,q_k}$ can be obtained by concatenation recursively $M^r_{q_{r-1}+1,q_r}$, $1\le r\le k$.
We use the superscript $w$ to represent the intermediates (recursively concatenating from right to left):
$$
R_{q_{w-1}+1,q_k}^w=\bigcup_{w\le r \le k}M^r_{q_{r-1}+1,q_r}. 
$$
Clearly we have the following bipartition:
$$
R_{q_{w-1}+1,q_k}^w= M^w_{q_{w-1}+1,q_w} \bigcup R_{q_{w}+1,q_k}^{w+1}.
$$
This decomposition of secondary structures allows us to compute the partition function efficiently.
%Thus, successive arc-removals decompose a structure into loops, each of which corresponding to a unique base-pair.  

%Define object, what is the partition function with filtration

%%%
%%%%%%%%%%%%%%%%%%%%%%%%%%%%%%%%%%%%%%%%%%%%%%%%%%%%%%%%%%%%%%%%%%%%%%%%%%%%%%%%%%%%%%%%%%%%%%%%%%%%%%%%%%
%%%
\begin{definition}
Given a structure $S$ and a reference sequence $\overline{\sigma}$, the partition function of $S$ 
with Hamming distance filtration $h$ to $\overline{\sigma}$ is given by
$$
Q^{S,\overline{\sigma}}_h = \sum_{\sigma, d(\sigma, \overline{\sigma}) = h} e^{\frac{-\eta(\sigma,S)}{RT}}, 
$$
where $\eta(\sigma, S)$ is the energy of $S$ on $\sigma$, $d(\sigma, \overline{\sigma})$ denotes the Hamming distance 
between $\sigma$ and $\overline{\sigma}$, $R$ is the universal gas constant and $T$ is the temperature. 
\end{definition}
%%%
%%%%%%%%%%%%%%%%%%%%%%%%%%%%%%%%%%%%%%%%%%%%%%%%%%%%%%%%%%%%%%%%%%%%%%%%%%%%%%%%%%%%%%%%%%%%%%%%%%%%%%%%%%
%%%

%How to compute 
%arc removal, partial order, base case 
%recursion, loop removal, junction base-pair, distribute d  
In the following we omit the explicit reference to $\overline{\sigma}$ and simply write $Q^S$. We shall
compute $Q^{S}_h$ following the secondary structure decomposition recursively. 
Suppose $(i,j)$ is a base-pair with its induced substructure $S_{i,j}$. Since the specific nucleotide composition
of $(i,j)$ may be involved in energy calculation of more than one loop, we introduce the partition functions of
substructures $X_{a,b}$, $Q^{X_{a,b}}_h(N_a,N_b)$, where $X_{a,b} = S_{a,b}$, $R^w_{a,b}$ or $M_{a,b}^w$, whose left and right endpoints $\sigma_a=N_a$ and $\sigma_b=N_b$ are determined and contributes 
$h$ to Hamming distance. We consider the set of subsequences 
$$
\{\sigma_{a,b}\in \mathcal{N}^{b-a+1}\mid d(\sigma_{a, b}, \overline{\sigma}_{a,b}) = h, \sigma_a= N_a, \sigma_b = N_b\},
$$
to which we refer to as $\mathcal{S}_h^{a,b}(N_a, N_b)$. Summing over all $\sigma_{a, b} \in \mathcal{S}_h^{a,b}(N_a, N_b)$
we derive
\begin{equation}\label{E:xx}
  Q_h^{X_{a,b}} (N_a, N_b) = \sum_{\sigma_{a, b} \in \mathcal{S}_h^{a,b}(N_a, N_b)}
      e^{\frac{-\eta(\sigma_{a,b}, X_{a,b})}{RT}}, 
\end{equation}
where $N_a, N_b \in \mathcal{N}$, $\mathcal{N}  = \{\bf A, U, C, G\}$. 

%hairpin
We next derive the recursion for $Q_h^{S_{i,j}} (N_i, N_j)$, computed from bottom to top.

{\bf Case 1:} $(i,j)$ is $\prec$-minimal, i.e.,~$S_{i,j}$ is a hairpin loop ($k=0$). 
By eq.~(\ref{E:xx}), summing over all subsequence $\sigma_{i,j}\in \mathcal{S} ^{i,j}_h(N_i, N_j)$ we derive
\begin{equation*}
Q_h^{S_{i,j}} (N_i, N_j)= 
 \sum_{\sigma_{i, j} \in \mathcal{S}_h^{i,j}(N_i, N_j)} e^{\frac{-\eta(\sigma_{i,j},S_{i,j})}{RT}}. 
\end{equation*}

{\bf Case 2:} $(i,j)$ is non-minimal and $k=1$, i.e.,~$L$ is an interior loop. Removing $(i,j)$ produces a single
$S_{p, q}$ as well as two intervals $[i+1, p-1]$ and $[q+1,j-1]$, either of which being possibly empty.
% if $p-1=i+1$ or $j-1=q+1$. 
Suppose $d(\sigma_{i,j}, \overline{\sigma}_{i,j}) = h$ and $d(\sigma_{p,q}, \overline{\sigma}_{p,q}) = t$, where
$0 \le t \le h$. Then the distance contribution from the intervals $[i, p-1]$ and $[q+1,j]$, $t_1$ and $t_2$,
satisfies $t_1+t_2=h-t$. Then $Q_h^{S_{i,j}} (N_i, N_j)$ equals 
\begin{align*}
\sum_{t, t_1, t_2} \sum_{N_p, N_q}  
\sum_{\sigma_{i,p} }
\sum_{\sigma_{q,j} }
%Q_h^{S_{i,j}} (N_i, N_j)= 
%\sum 
e^{\frac{-\eta(\sigma_{i,j}, L)}{RT}} Q_{t}^{S_{p, q}} (N_{p}, N_{q}), 
\end{align*}
where $t+t_1+t_2=h$, $N_p, N_q\in \mathcal{N}$,$\sigma_{i,p} \in \mathcal{S}_{t_1+\delta_p}^{i,p}(N_i, N_p)$ and $\sigma_{q,j} \in \mathcal{S}_{t_2+\delta_q}^{q,j}(N_q, N_j)$. Here 
$\delta_x = 1$ if $N_x = \overline{\sigma}_x$, and $\delta_x=0$, otherwise, for $x=p,q$. 

{\bf Case 3:} $(i,j)$ is non-minimal and $k\ge 2$, i.e.,~$L$ is a multiloop. In this case (in difference to the interior loops
analyzed above) the Turner energy model allows us to further decompose the energy of $\eta(\sigma,L)$ into independent
components, which in turn allows us to compute $Q_h^{S_{i,j}} (N_i, N_j)$ via recursive bipartitioning.
Removing $(i,j)$ produces $R_{q_0+1,q_k}^1$ as well as $[q_k+1, j-1]$, see Fig.~\ref{F:ham_decom} (A). The energy $\eta(\sigma_{i,j}, S_{i,j})$
is then given by
$$
\eta(\sigma, R_{q_0+1,q_k}^1)  + \alpha_{\text{mul}}+\eta_{\text{mul}} ((i,j))+\eta_{\text{mul}} ([q_k+1, j-1]),
$$
where $\alpha_{\text{mul}}$ is the energy contribution of forming a multiloop, $\eta_{\text{mul}} ((i,j))$ is the energy
contribution of base-pair $(i,j)$ in a multiloop, and $\eta_{\text{mul}} ([q_k+1, j-1])$ is the energy contribution
from the unpaired base interval in a multiloop.
The sum of the latter three component is denoted by $\eta^0$. 

Suppose 
%the Hamming distance contribution of base-pair $(i,j)$ being specifically $N_i$ paired with $N_j$ is $t_0$,
$d(\sigma_{q_0+1,q_k}, \overline{\sigma}_{q_0+1,q_k}) = t$ and $d(\sigma_{i,j}, \overline{\sigma}_{i,j}) = h$. Then the distance
contribution from the unpaired interval $[q_k+1, j-1]$ is $h-t-\delta_i-\delta_j$. 
%Therefore, going through all possible subsequences $\sigma_{q_k+1, j} \in \mathcal{S}^{\overline{\sigma}_{q_k+1, j}}_{h-t - \delta_i} (N_{q_k+1}, N_j)$,
%where $\delta_i = 1$ if $N_i=\overline{\sigma}_i$ otherwise $\delta_i = 0$, $0\le t \le h$, as well as all possible $N_{q_0+1}, N_{q_k} \in \mathcal{N}$,
Then $Q_h^{S_{i,j}}(N_i, N_j)$ equals
\begin{align*}
\sum_{t} \sum_{N_{q_0+1}, N_{q_k}}
\sum_{\sigma_{q_k, j} } 
%Q_h^{S_{i,j}}(N_i, N_j)=
%\sum
e^{\frac{-\eta^0}{RT}} Q_{t}^{R_{q_0+1,q_k}^1}(N_{q_0+1}, N_{q_k}), 
\end{align*}
where $0\le t \le h$, $N_{q_0+1}, N_{q_k} \in \mathcal{N}$ and $\sigma_{q_k, j} \in \mathcal{S}_{h-t-\delta_i+\delta_{q_k}}^{q_k, j}(N_{q_k}, N_j)$.

This brings us to substructures $R_{q_{w-1}+1,q_k}^w$, $1\le w \le k$, which decompose into (are concatenations of)
$M^w_{q_{w-1}+1,q_w}$ and $R^{w+1}_{q_w+1, q_k}$. 
For notational convenience we set $r = q_{w-1}+1$, $s=q_w$, $u = q_w+1$ and $v= q_k$. 
Suppose $d(\sigma_{r, s}, \overline{\sigma}_{r, s}) = t$ 
and $d(\sigma_{r, v}, \overline{\sigma}_{r, v}) =h$, then $d(\sigma_{u, v}, \overline{\sigma}_{u, v}) = h-t$. 
We obtain for $Q_h^{R^w_{r, v}} (N_{r}, N_{v})$ the expression 
\begin{align*}
\sum_{t}\sum_{N_{s}, N_{u}}
%\sum
Q_{t}^{M_{r, s}^w} (N_{r}, N_{s})
Q_{h-t}^{ R_{u, v}^{w+1}} (N_{u}, N_{v}), 
\end{align*}
where  $0\le  t\le h$ and  $N_{s}, N_{u} \in \mathcal{N}$.

The substructures $M_{q_{w-1}+1,q_w}^w$ are concatenations of $[q_{w-1}+1, p_w-1]$ and $S_{p_w,q_w}$, for $1\le w \le k$.
For notational convenience we set $r = q_{w-1}+1$, $s=p_w-1$, $u = p_w$ and $v= q_w$. 

Suppose $d(\sigma_{u, v}, \overline{\sigma}_{u, v}) = t$ and $d(\sigma_{r, v}, \overline{\sigma}_{r, v})
= h$, then the Hamming distance of $[r, s]$ to the corresponding $\overline{\sigma}$-interval is $h-t$.

Summing over $0\le t \le h$, all $N_{p_w-1}, N_{p_w} \in \mathcal{N}$, all $\sigma_{q_{w-1}+1, p_w-1} \in 
\mathcal{S}_{h-t}^{q_{w-1}+1, p_w-1}(N_{q_{w-1}+1}, N_{p_w-1})$, we derive for $Q_{h}^{M_{r,v}^w} (N_{r}, N_{v})$ 
\begin{align*}
\sum_{t} \sum_{N_{s}, N_{u}} \sum_{\sigma_{r, s}}
%\sum
Q_{h-t}^{S_{u, v}} (N_{u}, N_{v}) 
%e^{\frac{-\eta_{\text{mul}}((p_w, q_w))-\eta_{\text{mul}}([q_{w-1}+1, p_w-1]) }{RT}}. 
e^{\frac{-\eta^w}{RT}}, 
\end{align*}
where $0\le t \le h$, $N_{s}, N_{u} \in \mathcal{N}$, $\sigma_{r, s} \in \mathcal{S}_{h-t}^{r, s}(N_{r}, N_{s})$
and $\eta^w = \eta_{\text{mul}}((u, v))+\eta_{\text{mul}}([r, s])$. 
Here $\eta_{\text{mul}}((u, v))$ is the energy contribution of base-pair $(u, v)$ in a multiloop and 
$\eta_{\text{mul}}([r, s])$ is the contribution of segment of unpaired bases in a multiloop.
We present the recursions in Fig.~\ref{F:ham_decom} (B).

The introduction of the intermediate substructures $M_{q_{w-1}+1,q_w}^w$ and $R^w_{q_{w-1}+1,q_k}$ avoids processing concatenation
of substructures simultaneously, which would result in a $O(h^{k-1})$ time complexity.
The family of intermediate substructures $M_{q_{w-1}+1,q_w}^w$ and $R^w_{q_{w-1}+1,q_k}$ 
remedies this problem by executing one concatenation at each step, effectively bipartitioning and requiring a time
complexity of $O(h)$. In total we encounter $k-1$ such bipartition, resulting in a $(k-1)O(h)$ time complexity.
Since there are $O(n)$ base-pairs in a structure and each entails to compute $O(h)$ partition functions, we have
to consider $O(h n)$ partition functions. As a result the time complexity of the algorithms is $O(h^2 n)$.

Following this recursion, $Q_h^{S_{i,j}}(N_i, N_j)$ can be computed from bottom to top as claimed.
The recursion terminates, when reaching the rainbow, $(0,n+1)$. The partition function of $S$ with Hamming 
distance filtration $h$ to $\overline{\sigma}$ is given by $Q_h^S = Q_h^{S_{0,n+1}}(N_0, N_{n+1})$, where 
$N_0$ and $N_{n+1}$ are ``formal'' nucleotides, discussed above.

Having computed the partition function $Q_h^{S,\overline{\sigma}}$, we implement the Boltzmann sampler of RNA
sequences having a fixed Hamming distance $h$ to $\overline{\sigma}$ from $S$ following the classical stochastic
backtracking method introduced by \citep{Ding:03}. This process first samples the nucleotides in the exterior
loop and then subsequently samples the loops following the partial order $\prec$ from top to bottom until
reaching the hairpin loops.

Since the time complexity of computing a loop energy in Turner's model is constant, 
the worst case time complexity of the sampling process is $O(n^2)$ \citep{Ding:03}, and applying the Boustrophedon
technique for Boltzmann sampling, introduced in \citep{Ponty:08,Nebel:11}, reduces the time complexity to
$O(n\log n)$ on average.

%%%
%%%%%%%%%%%%%%%%%%%%%%%%%%%%%%%%%%%%%%%%%%%%%%%%%%%%%%%%%%%%%%%%%%%%%%%%%%%%%%%%%%%%%%%%%%%%%%%%%%%%%%%%%%%
%%%

\section{Results}\label{S:Result}

%%%
%%%%%%%%%%%%%%%%%%%%%%%%%%%%%%%%%%%%%%%%%%%%%%%%%%%%%%%%%%%%%%%%%%%%%%%%%%%%%%%%%%%%%%%%%%%%%%%%%%%%%%%%%%%
%%%

In this section, we first study the Hamming distance distribution of sequences generated via the unrestricted dual
sampler \cite{Barrett:17}. We perform this analysis for natural sequences as well as random sequences. The resulting distribution
shows that a simple rejection sampler is rather inefficient and motivates the algorithm derived in Section~\ref{S:Method}.

Here we apply the refined Boltzmann sampler in order to gain deeper insight into IFR and neutral paths.
First, given a sequence-structure pair, $(\overline{\sigma},S)$ we study the rate at which sampled sequences, filtered by
Hamming distance, fold into $S$. Secondly, we apply the sampler in order to develop an efficient heuristic that constructs paths
within neutral networks, i.e.,~given two sequences, both of which folding into a fixed structure $S$, we identify a path consisting
of sequences all of which folding into $S$, such that two consecutive sequences on the path differ only by a point- or pair-mutations.

%%%
%%%%%%%%%%%%%%%%%%%%%%%%%%%%%%%%%%%%%%%%%%%%%%%%%%%%%%%%%%%%%%%%%%%%%%%%%%%%%%%%%%%%%%%%%%%%%%%%%%%%%%%%%%%%%%
%%%

\begin{figure}[h]
\begin{center}
\includegraphics[width=0.8\columnwidth]{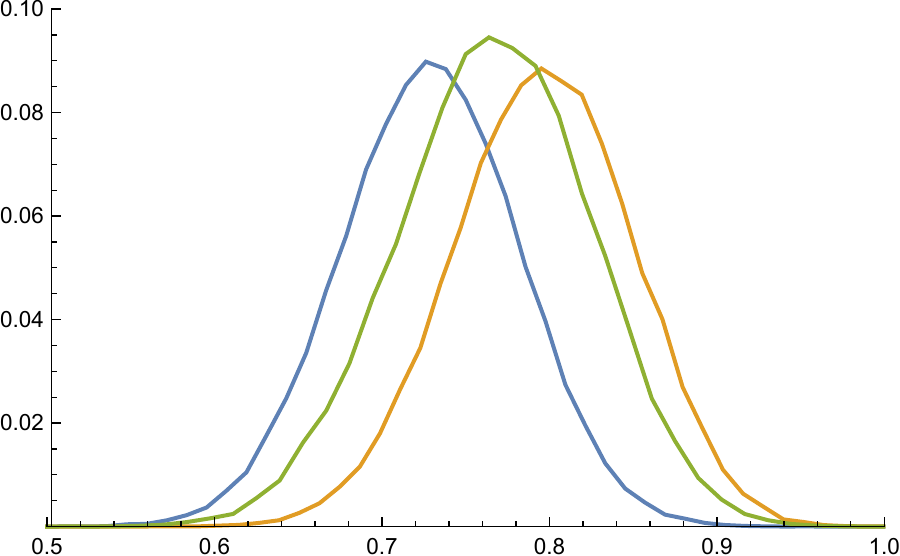}
\end{center}
\caption{\small Hamming distance distribution of sampled sequences for three sequence-structure pair of the human microRNA
  let-7 family (hum01, hum09, and hum10). For each pair we sample $5\times 10^4$ sequences, using the unrestricted sampler in
  \citep{Barrett:17}. We display the Hamming distance distribution of the sampled sequences to the natural sequence. 
The $x$-axis is the Hamming distance normalized by the sequence length, and the $y$-axis is frequency sampled sequences. 
} 
\label{F:distribution}
\end{figure}

{\bf Hamming distance distribution:}
we consider $12$ sequence-structure pairs from the human microRNA let-7 family in miRBase \citep{kozomara2013mirbase}. For each pair we
sample $5 \times 10^4$ sequences using the unrestricted sequences sampler in \citep{Barrett:17}. Then we compute the Hamming distance
distribution to the natural sequence of the sampled sequences. The distances are normalized by sequence length. We display in
Fig.~\ref{F:distribution} the distance distribution of three distinguished sequence-structure pairs, whose mean distance is in some
sense minimal, typical, and maximal, respectively. The full spectrum of these distributions is presented in the SM, Fig.~1. 

The data show that the sampled sequences have distances between $60\%$ and $90\%$ of the sequence length to the 
reference sequence. The mean distance is $70\%$ to $80\%$ of the sequence length, indicating that the unrestricted
sampler in \citep{Barrett:17} does not produce the full spectrum of distance classes. 
To analyze the robustness of these findings, we replace the natural sequence by $10$ random sequences folding into
the reference structure and compute the distance distribution for the pairs hum05 and hum09, see SM Fig.~2 and Fig.~3.
We observe that most of the sampled sequences still have a Hamming distance of $60\%$ to $90\%$ of the sequence length.
However, we observe some variations of the distribution, whence we have not only a dependence on structure, but also on
the reference sequence. This implies the sampled sequences are not uniformly distributed. More interestingly, we observe
that the Hamming distance distribution corresponds to the natural sequence is the least concentrated around the mean.

%%%
%%%%%%%%%%%%%%%%%%%%%%%%%%%%%%%%%%%%%%%%%%%%%%%%%%%%%%%%%%%%%%%%%%%%%%%%%%%%%%%%%%%%%%%%%%%%%%%%%%%%%%%%%%%%%%
%%%

{\bf Inverse fold rate:} we study now the inverse fold rate (IFR) of the sampled sequences with respect to different
sequence-structure pairs, for different Hamming distances. We associate an indicator variable to each sampled sequence: taking
the state $1$ if the sequence actually folds into the reference structure and $0$, otherwise. By construction the IFR is the mean of this
random variable and we consider the IFR of a sequence-structure pair as a function of the Hamming distance, $h$, to the
reference sequence, $\text{IFR}(h)$. 

Given a sequence-structure pair $(\overline{\sigma}, S)$, we sample $5\times 10^4$ sequences from $S$ having a fixed 
Hamming distance, $h$, where $h$ is ranging from $1$ to $20$. 
Then $\text{IFR}(h) = U/M$ where $U$ is the number of sampled sequences folding back to $S$ and $M$ is the 
sample size. 
%$$
%\text{IFR}(h) = \frac{\text{\# sampled sequences fold back to $S$}}{\text{Sample size}}. 
%$$

\begin{figure}[h]
\begin{center}
\includegraphics[width=0.8\columnwidth]{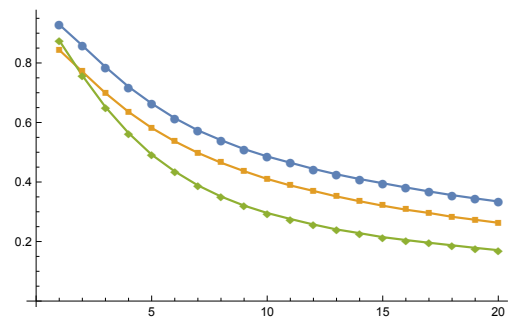}
\end{center}
\caption{\small Mean IFR of sequence-structure pairs of microRNA let-7 family of human (blue),
  lizard (yellow) and drosophila (green). The $x$-axis represents Hamming distance and the $y$-axis represents the IFR. 
} 
\label{F:IFR_mean}
\end{figure}

We consider the microRNA let-7 family of three species: human (hum01-12) , lizard (liz01-11) and drosophila (dro01-08),
computing their IFRs respectively. We display the mean\footnote{taken over the entire collection of let-7 micrRNAs of a
  given species in the database} IFR of the three species in Fig.~\ref{F:IFR_mean} and the IFR distributions of individual
pairs within the three species in the SM, Fig.~4, Fig.~5 and Fig.~6, respectively. 

\begin{figure}[h]
\begin{center}
\includegraphics[width=0.8\columnwidth]{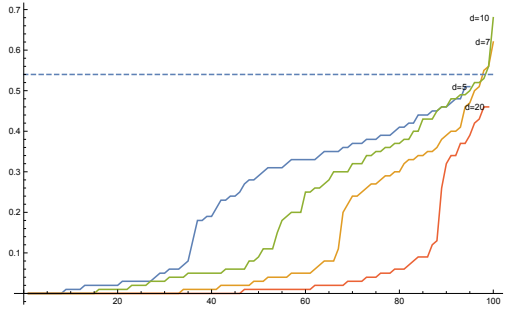}
\end{center}
\caption{\small $\text{IFR}(5)$ of the natural sequence-structure pair versus adjacent sampled sequences. We compare the natural
sequence-structure pair of hum04 and sample $100$ sequences of distance $5$, $7$, $10$ and $20$, respectively. 
We display $\text{IFR}(5)$ of the induced sequence-structure pairs sorted by their $\text{IFR}(5)$ in increasing order
for distance $5$ (blue), $7$ (yellow), $10$ (green) and $20$ (red). $\text{IFR}(5)$ of the natural sequence-structure pair is
displayed as the dashed line. Here the $x$-axis is labeled by the sorted sequence-structure pairs and the $y$-axis represents
the IFR.
} 
\label{F:neighborhoodIFR} 
\end{figure}

The Fig.~\ref{F:IFR_mean} shows that human has the highest, mean IFR while drosophila has the lowest. In addition, the mean IFR
decreases for human significantly slower than for drosophila. To analyze robustness and dependencies of these findings, we
compute the IFR of random sequence-structure pairs and to those of natural pairs. In the following we restrict ourselves to the
hum04 sequence-structure pair.
We first consider random sequences compatible with the hum04-structure and thereby create new sequence-structure pairs.
Then we compute $\text{IFR}(5)$ of these pairs by sampling $5\times 10^4$ sequences of Hamming distance $5$. The $\text{IFR}(5)$
is almost zero for these random sequences indicating that random compatible sequences have little or no connection with
the hum04-structure. To identify sequences that are closer related to the hum04-structure, we use our sampler, creating $100$
sequence-structure pairs by sampling sequences from the natural pairs of distance $5$, $7$, $10$ and $20$,
respectively. Then we combine the newly sampled sequences with the hum04-structure, creating new sequence-structure pairs.
We compute their $\text{IFR}(5)$ and sort the pairs by their $\text{IFR}(5)$ in increasing order, see Fig.~\ref{F:neighborhoodIFR}.
For reference purposes we display $\text{IFR}(5)$ of the natural sequence-structure pair as a dashed line.

Fig.~\ref{F:neighborhoodIFR} shows that $\text{IFR}(5)$ of the natural sequence-structure pair is above the $95$ percentile,
i.e.,~better than almost all of the newly created pairs. Furthermore, there exists very few pairs such that
$\text{IFR}(5) \in [0.1, 0.3]$ holds. The proportion of sequence having high  $\text{IFR}(5)$, i.e.,~$\text{IFR}(5) >0.3$
drops when the sampled sequence have higher Hamming distance. This finding suggests that the natural pair is locally optimal. 

%%%
%%%%%%%%%%%%%%%%%%%%%%%%%%%%%%%%%%%%%%%%%%%%%%%%%%%%%%%%%%%%%%%%%%%%%%%%%%%%%%%%%%%%%%%%%
%%%

{\bf Neutral paths:} as discussed in Section~\ref{S:Into}, connectivity is of central importance in neutral networks.
Combined with some form of density, it allows genotypes to explore, by means of point- or pair-mutations, extended portions of
sequence space. An exhaustive analysis of connectivity is not feasible even for relatively short sequence length, whence 
the explicit construction of specific paths within the neutral network is the best possible outcome.
To be clear, let us first specify the neutral path problem:

\emph{Given two sequences $\sigma_1$ and $\sigma_2$, both folding into the structure $S$, identify a path
  $\sigma_1=\tau_0,\tau_1, \ldots, \tau_k=\sigma_2$, such that\\
  (*) for all $\tau_i$, $0\leq i \leq k$, folds to $S$,\\
  (**) $\tau_{i+1}$ is obtained from $\tau_i$ by either a compatible point- or a base-pair mutation.}\\

The construction of such ``neutral paths'' has been studied in \citep{gobel2002rna} using a proof idea that facilitates the
construction of neutral paths, for fixed, finite distance $d$, in random induced subgraphs. However \citep{gobel2002rna}
exhaustively checks whether such paths are neutral or not, irrespective of $d$, a task that becomes impracticable for large
$d$. At present, there is no efficient way of finding neutral paths in a neutral networks induced by folding algorithms, in
particular in case of the distance between the two sequences being large. In the following we shall employ our sampling
algorithm in order to derive an efficient heuristic to solve the neutral path problem. 

  Certainly, given $\sigma_1$ and $\sigma_2$, both folding into $S$, one can always construct a path between them using the two above
  moves. By construction this is a $S$-compatible path.
  Furthermore, there exists a minimum number of moves that have to be performed to traverse from $\sigma_1$ to $\sigma_2$.
  We refer to this as the $S$-compatible distance between $\sigma_1$ and $\sigma_2$, $d_S(\sigma_1,\sigma_2)$.
  Clearly, we have $\frac{1}{2}d \leq d_S\leq d$, for any $S$-compatible sequences. A neutral path, whose length equals the
  $S$-compatible distance is called a shortest neutral path. In the context of the neutral path problem, we do not require the
  paths to be minimal in length.
  
 {\bf Case 1:} $d(\sigma_1,\sigma_2)\leq 5$. Here we exhaustively search all shortest $S$-compatible paths between $\sigma_1$
 and $\sigma_2$ and check for neutrality. Note that we always have $d_S\leq d$, thus in the worst case, we need to
 check $5!=120$ different paths and fold $2^5=32$ different sequences. This is feasible for sequence lengths shorter that
 $10^3$ nucleotides, using standard secondary structure folding algorithms \citep{Zuker:81, Hofacker:94a}. 

\begin{figure}[h]
\begin{center}
\includegraphics[width=0.8\columnwidth]{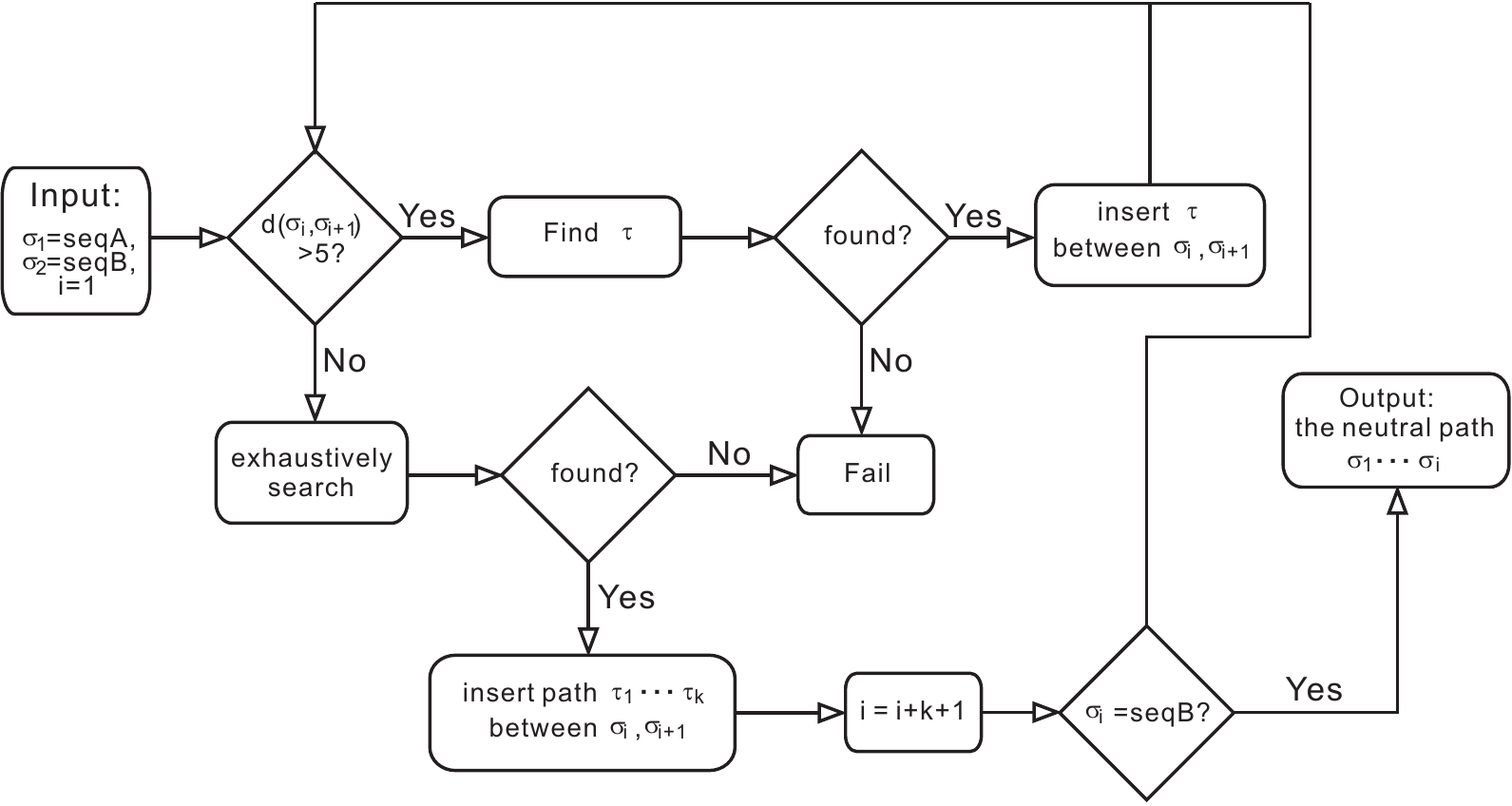}
\end{center}
\caption{\small The algorithm. 
} 
\label{F:flow}
\end{figure}

 {\bf Case 2:} $d(\sigma_1,\sigma_2)> 5$. Suppose $\sigma_1$ and $\sigma_2$ have Hamming distance $h$. We sample $m$ sequences
 from $\sigma_1$ with respect to $S$ with distance filtration $h/2$. $m=1000$ typically suffices but higher sampling size can
 easily be realized if the IFR is too low. We then select such a sequence with minimum Hamming distance to $\sigma_2$, denoted
 by $\tau_s$. We have $d(\sigma_1, \tau_s) = h/2 = h_1$ and $d(\tau_s, \sigma_2) = h_2$, where $h_1+h_2\ge h$.
 If $h_2>h$ we claim the process fails and we conclude we can not find a neutral path between $\sigma_1$ and $\sigma_2$.
 Otherwise, we repeat the process between $\sigma_1$ and $\tau_s$, and between $\tau_s$ and $\sigma_2$, differentiating
 Case 1 and Case 2. We show the flow of the algorithm in Fig.~\ref{F:flow}. 
 
\begin{figure}[h]
\begin{center}
\includegraphics[width=0.8\columnwidth]{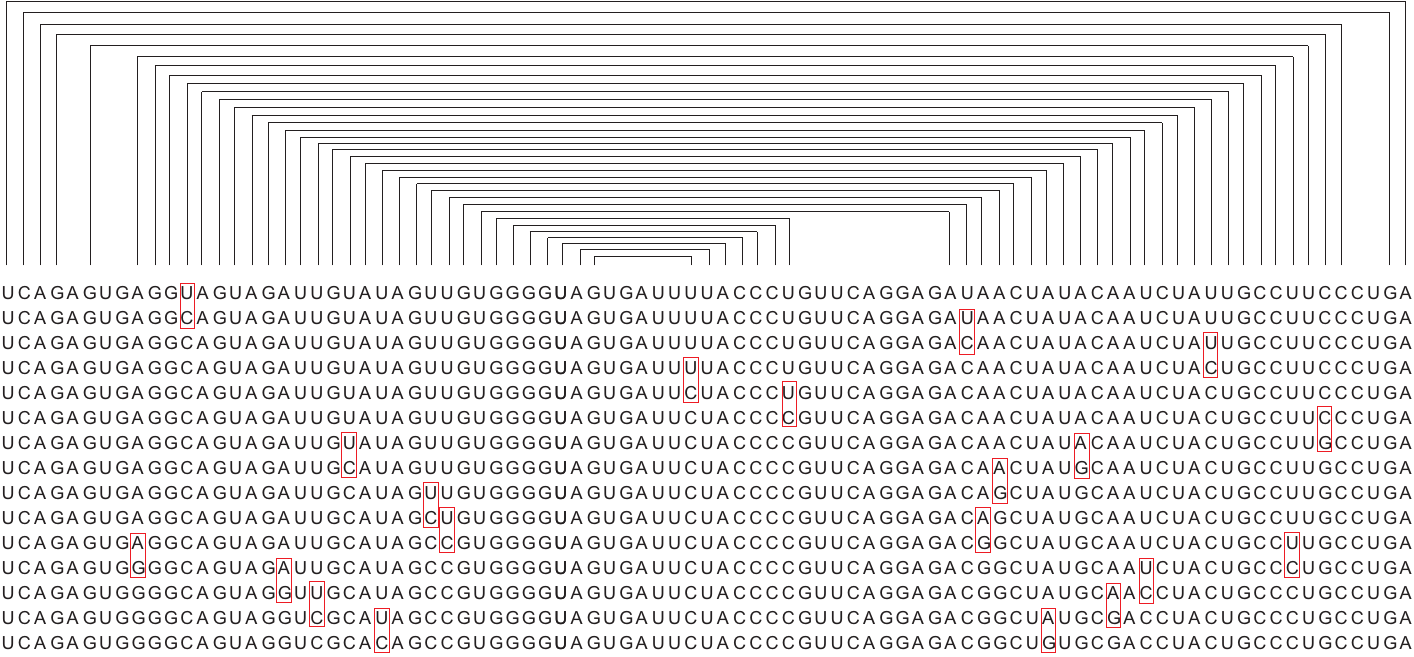}
\end{center}
\caption{\small A neutral path connecting the natural sequence of hum08 to a sequence having 
Hamming distance $20$. All sequences along the path fold into the natural structure of hum08. 
This particular path has length $14$ and consists of $8$ point- and $6$ base-pair mutants. 
} 
\label{F:Npath}
\end{figure}

 The process either fails at some point of the iteration or produces recursively a neutral path. We illustrate 
 a particular neutral path, connecting the natural sequence of hum08 to a Hamming distance $20$ sequence in Fig.~\ref{F:Npath}. 

As for algorithmic performance: for hum04 we consider the natural sequence and structure pair and sample $100$ sequences
of Hamming distance $20$, $19$ of which being neutral. We pair each of these with the natural sequence and compute a
neutral path. The algorithm succeeded $18$ times and failed to produce a neutral path once.
For hum08 we perform the same experiment for Hamming distances $20$ and $40$, respectively.
In case of Hamming distance $20$ we find neutral $85$ sequences, for these the algorithm succeeds $83$ times and fails twice.
For distance $40$ we find $53$ neutral sequences: the algorithm succeeds $49$ times and fails in four instances. 

For a low level organism microRNA, bra01, which is a Branchiostoma micro RNA, at distance $20$, we find $22$ neutral
sequences: $16$ successes and $6$ fails.

%%%
%%%%%%%%%%%%%%%%%%%%%%%%%%%%%%%%%%%%%%%%%%%%%%%%%%%%%%%%%%%%%%%%%%%%%%%%%%%%%%%%%%%%%%%%%%%%%%%%%%%%%%%%%%%%%%%%%%%%%
%%%

\section{Discussion}\label{S:D}

%%%
%%%%%%%%%%%%%%%%%%%%%%%%%%%%%%%%%%%%%%%%%%%%%%%%%%%%%%%%%%%%%%%%%%%%%%%%%%%%%%%%%%%%%%%%%%%%%%%%%%%%%%%%%%%%%%%%%%%%%
%%%

%What does the sampler do?
The problem of finding a sequence that folds into a given structure, $S$, has first been studied in \citep{Hofacker:94a}.
The algorithm consists of two parts: first it constructs a random $S$-compatible sequence and secondly it performs an
adaptive walks of point mutant in the sequence such that facilitates identifying a sequence that folds into $S$.
In this process, neither an inverse fold solution is guaranteed nor the number of adaptive walks required is understood.
\citep{Busch:06} shows that such adaptive walks can be constructed much more easily, when proper care is taken where
the process actually initiates. Namely, choosing the $S$-compatible sequence such that it minimizes the free energy with
respect to $S$. \citep{levin2012global,Clote:16} observe that Boltzmann sampled sequences exhibit a distinctively higher
rate of folding again into $S$.

The high IFR of sampled sequences from a structural ensemble is not only useful in finding candidate sequences for
inverse folding problems reflects in some sense the robustness of the structure. High IFR in structural ensembles
indicate that the structure is likely preserved within limited energy change and mutations on a sequence. This is
quite subtle as competing structural configurations may offer a fixed sequence an even lower and thus more preferable
free energy. The problem can therefore not be reduced to minimizing free energy of sequences with respect to a fixed
structure, it is context dependent.

However, sampled sequences from the structural ensemble are not conserved and differ vastly from each other. It is natural to
bring evolutionary trajectories into the picture, necessitating the ability to study Boltzmann sampled sequences having fixed
Hamming distance to some reference sequence. This allows us to investigate local features and brings sequence information into
the picture. By introducing the Hamming distance filtration, we can zoom into a specific sequence as well as its neighborhood
in the structural ensemble. These sequences are not only sorted by the given structure but also evolutionary close to the
reference sequence. This approach shifts focus to considering sequences and structures as pairs, as discussed in 
\citep{Barrett:17}

%What is the algorithm intellectual merit?
\citep{levin2012global} presents a Boltzmann sampler of sequences from a structural ensemble with Hamming distance restriction.
The algorithm described in \citep{levin2012global} constitutes a constrained version of the algorithm described in
\citep{waldispuhl2008efficient}, having a time complexity of $O(h^2n^3)$ where $h$ is the Hamming distance.
The partition function of sequences with distance filtration on all secondary structures is computed, requiring to consider
all subintervals of $[1,n]$ as well as an additional for-loop index, induced by the concatenation of two substructures.

Our algorithm has a time complexity of $O(h^2 n)$, a result of different recursions. We utilize the hierarchical
organization, or equivalently the induced partial order of the arcs of a secondary structure structure, together with the fact
that free energy is computed based on loops. This allows us to compute the partition function from the inside to the outside
(bottom to top from the tree prospective). The routine is purely driven by the fixed structure, whence no redundant information
is computed.

The dual sampler, i.e.,~the Boltzmann sampler of sequences with respect to a fixed structure, with Hamming distance filtration
(enhanced sampler) brings sequence information into the picture. This enables us to study evolutionary questions with the
enhanced dual sampler. Inverse fold rates and their Hamming distance dependence but also questions as the structural diversity of
the derived sequences can be analyzed effectively with the enhanced sampler. These studies follow the generalized scheme of
inferring information on any random variable over sequences partitioned into Hamming classes. Hamming classes in this sense can
be viewed as blocks of a partition to which a random variable can be restricted to.
Our analysis of IFR gives first indications that microRNAs of highly evolved organisms exhibit higher
robustness than those of organisms of lower level: in the context of evolutionary optimization achieving
robustness of evolved phenotypes is an advancement.

The enhanced sampler is furthermore useful for construction neutral paths. The naive approach to identifying neutral paths
between two given sequences $\sigma_1$ and $\sigma_2$ \citep{gobel2002rna} is to exhaustively check all shortest compatible paths
between them for neutrality. While this is feasible for small $d_S$ is small, as $d_S$ increases, the number of these shortest
paths grows hyper-exponential. In addition a neutral path might still exist even when all shortest compatible paths are not
neutral.
The enhanced sampler shows that even at large Hamming distance, sampled sequence have a high inverse fold rate, provided
reference sequence and structure are natural. This motivated the ``divide and conquer'' strategy employed to construct the
neutral paths.  We use he enhanced sampler to construct recursively ``intermediate'' sequences, that are traversed by the
neutral path. Iterating this process, we can reduce the Hamming distances to the point where exhaustive search becomes
feasible, see Fig.~\ref{F:flow} in Section~\ref{S:Result}.

It is possible that the shortest possible neutral path has length strictly greater than the $S$-compatible distance, however,
Case 1 does not consider any such paths. In order to validate the approach of Case 1, we consider sequence-structure pairs of
the microRNA let-7 family across various species (human, cattle lizard and other low level organism, $12$ pairs for each class)
as the origin. Then for each sequence-structure pair, we identify inverse fold solutions by dual sampling $1\times 10^4$
sequences of Hamming distance $5$ and consider all neutral solutions\footnote{on average $5\times 10^3$ neutral sequences were found}
as the terminus. By exhaustive search, we observe that for all of these sequence pairs, there exists s neutral path, whose length
is equal to the $S$-compatible distance.

%Energy distribution of sampled sequences without filtration. Natural structures of hsa 00067, 000063,  bta 052509. 
%Purpose of this figure: XX.

\section{Acknowledgments}
We want to thank Stanley Hefta and Peter Stadler for their input on this manuscript. 
We gratefully acknowledge the help of Kevin Shinpaugh and the computational support team at BI, 
Mia Shu, Thomas Li, Henning Mortveit, Madhav Marathe and Reza Rezazadegan for discussions.
The fourth author is a Thermo Fisher Scientific Fellow in Advanced Systems for Information Biology and
acknowledges their support of this work.

\bibliographystyle{natbib}
\bibliography{sample_ham_result}

\end{document}